# Chemical modifications and stability of phosphorene with impurities:

# A first principles study

D. W. Boukhvalov,[1,*] A. N. Rudenko,[2] D. A. Prishchenko,[3] V.G. Mazurenko,[3] M. I. Katsnelson[2,3]

[1] *Department of Chemistry, Hanyang University, 17 Haengdang-dong, Seongdong-gu, Seoul 133-791, Korea*
[2] *Institute for Molecules and Materials, Radboud University, Heijendaalseweg 135, 6525 AJ Nijmegen, The Netherlands*
[3] *Theoretical Physics and Applied Mathematics Department, Ural Federal University, Mira Str. 19, 620002 Ekaterinburg, Russia*

*We perform a systematic first-principles study of phosphorene in the presence of typical monovalent (hydrogen, fluorine) and divalent (oxygen) impurities. The results of our modeling suggest a decomposition of phosphorene into weakly bonded one-dimensional (1D) chains upon single- and double-side hydrogenation and fluorination. In spite of a sizable quasiparticle band gap (2.29 eV), fully hydrogenated phosphorene found to be dynamically unstable. In contrast, full fluorination of phosphorene gives rise to a stable structure, being an indirect gap semiconductor with the band gap of 2.27 eV. We also show that fluorination of phosphorene from the gas phase is significantly more likely than hydrogenation due to the relatively low energy barrier for the dissociative adsorption of $F_2$ (0.19 eV) compared to $H_2$ (2.54 eV). At low concentrations, monovalent impurities tend to form regular atomic rows phosphorene, though such patterns do not seem to be easily achievable due to high migration barriers (1.09 and 2.81 eV for $H_2$ and $F_2$, respectively). Oxidation of phosphorene is shown to be a qualitatively different process. Particularly, we observe instability of phosphorene upon oxidation, leading to the formation of disordered amorphous-like structures at high concentrations of impurities.*

*E-mail: danil@hanyang.ac.kr

**1. Introduction**

Recent success in isolation of ultrathin layers of black phosphorus has raised interest for its electronic properties.[1-10] The presence of an appreciable direct band gap[11,12] and high carrier mobility[3-7,9] make few-layer black phosphorus a promising candidate for novel semiconductor applications. At the same time, nontypical attributes of this two- dimensional (2D) material such as strong anisotropy of the electronic properties and their pronounced thickness-dependence provide new avenues for theoretical investigations.[11-15] From the point of view of practical applications, the influence of environment plays a crucial role in the performance of potential 2D electronic devices. Unlike other known 2D materials, few-layer black phosphorus samples demonstrate high reactivity, resulting in a fast degradation of the crystals under ambient conditions,[7,8,16-18] which will inevitably hinder the performance of potential phosphorene devices.[19] Existing theoretical studies point to a strong affinity of black phosphorus toward ubiquitous molecules such as water[8] and oxygen,[20-22] as well as to a strong coupling with monoatomic impurities (such as, e.g., hydrogen, fluorine, and oxygen)[23-25] and relatively low vacancy formation energies.[26-28] At the same time, the detailed chemical mechanism and thermodynamic conditions leading to the degradation of black phosphorus are currently unknown. In particular, stability of defective phosphorene remains largely unclear. These issues motivated us to systematically explore chemical properties of phosphorene (black phosphorus monolayer) in the presence of typical monovalent and divalent impurities.

Apart from the negative consequences due to the interaction with impurities, a controlled passivation of black phosphorus might lead to new stable structures as it shows the prominent example of graphene derivatives.[29-31] Last not least, the chemical reactivity of black phosphorus might be exploited for the purposes of gas sensing due to the reported adsorption selectivity.[22,32] Recent theoretical works discussed chemical properties of phosphorene in the presence of hydrogen,[23,24] fluorine,[23] and oxygen.[20,25,33] In these works, however, the authors discussed either certain single impurities in phosphorene or its stoichiometric derivatives, whereas the mechanism of a step-by-step functionalization is not yet understood. At the same time, the experience of the modeling of graphene chemical modifications demonstrates the need for a step-by-step description of the functionalization process because the existence of desired atomic structures may turn out to be unlikely due to the presence of highly unfavorable intermediate states[34] or high energy barriers of the molecular dissociation upon adsorption. Another important aspect of chemical functionalization is the dynamical stability of the reaction products of phosphorene and, particularly, its derivatives. This issue, to the best of our knowledge, has previously been partially addressed only in the context of oxygen impurities.[25,33]

Here, we perform a first-principles modeling of phosphorene in the presence of hydrogen, fluorine and oxygen impurities. We first consider the limit of maximum surface coverage (one impurity per phosphorus atom) by hydrogen or fluorine atoms. In this case, the phosphorene structure splits into weakly bound 1D chains arranged in the zigzag direction. We then analyze electronic properties and dynamical stability of fully hydrogenated and fluorinated phosphorene structures by calculating the quasiparticle band structures and vibrational spectra for the corresponding compounds. Both systems are found to be indirect gap semiconductors with an energy gap of ~2 eV. However, the hydrogenated structure does exhibit soft modes in the vibrational spectrum, which suggests its instability with respect to further structural transformation. On the other hand, fully fluorinated phosphorene is found to be dynamically stable. As a next step, we consider atom-by-atom single- and double-side functionalization of phosphorene and find the most favorable adsorption patterns for different impurity concentrations. We find that those correspond to a regular surface distribution of hydrogen and fluorine impurities, representing one-dimensional (1D) rows in the zigzag direction of phosphorene. Such regular patterns, however, does not seem to be directly observable from the adsorption of molecular impurities ($H_2$, $F_2$) due to high energy barriers governed by molecular dissociation and atomic migration. As a special case, we also consider oxygen impurities on phosphorene and show that they behave qualitatively different compared to hydrogen and fluorine. Particularly, oxygen impurities tend to form an irregular arrangement of atoms on the phosphorene surface, while at high concentrations a transition to a disordered amorphous-like structures is observed.

## 2. Computational methods

We used the pseudo-potential code SIESTA[35] to perform energy calculations of the various atomic structures of functionalized phosphorene within the density functional theory (DFT). All calculations were carried out by the local density approximation (LDA)[36] with spin polarization. To model the phosphorene monolayer, we used a rectangular supercell with 72 phosphorus atoms (see Fig. 1a). The atomic positions were fully optimized. During the optimization, the ion cores are described by norm-conserving non-relativistic pseudo-potentials[37] with cut off radii 1.85, 1.15, 1.14 and 1.25 a.u. for P, O, F and H, respectively. The wave functions were expanded with a double-ζ plus polarization basis of localized orbitals for all atoms except hydrogen and double-ζ for hydrogen. The force and total energy was optimized with an accuracy of 0.04 eV/Å and 1 meV, respectively. All calculations were performed with an energy mesh cut-off of 360 Ry and a **k**-point mesh of 12×10×1 in the Monkhorst-Pack scheme.[38] The formation energies for the functionalization were calculated using a standard formula: $E_{form} = (E_{host+guest} - [E_{host} + nE_{guest}])/n$, where $E_{host}$ is the total energy of the system before the adsorption of $n$ atoms, $E_{guest}$ is the total energy per atom of molecular oxygen (in the triplet spin-polarized state), hydrogen or fluorine in an empty box.

The band structures of fully hydrogenated and fluorinated phosphorene have been calculated by the $G_0W_0$ method within the projected augmented-wave (PAW) formalism as implemented in the Vienna *ab initio* simulation package (VASP).[39-41] To this end, a primitive cell containing 4 phosphorus and 4 hydrogen atoms relaxed at the LDA level was used. As a starting point for $G_0W_0$ calculations, we used the LDA wave functions, for which an energy cutoff of 250 eV and the convergence criterion of $10^{-8}$ eV were employed. To achieve numerical accuracy in $G_0W_0$ calculations, the number of virtual orbitals was chosen to be ~35 times greater than the number of occupied bands, whereas for integration along the frequency axis 70 grid points were used. The Brillouin zone was sampled by a (12x10x1) **k**-point mesh. The phonon calculations were performed by means of the frozen phonon method as implemented in the PHONOPY[42] code. The force constants were derived from forces calculated within the PAW method on a 4x3x1 supercell by atomic displacements of 0.01 Å along the lattice vectors.

## 3. Structure, electronic properties and stability of single- and double-side hydrogenated and fluorinated phosphorene

We first discuss general mechanism of the bond formation in phosphorene with monovalent impurities. In case of graphene, carbon atoms are $sp^2$ hybridized forming covalent bonds with three in-plane σ-orbitals and one $p_z$-orbital being perpendicular to the graphene plane. Chemisorption of species on graphene results in the $sp^2$ to $sp^3$ transition, where a π-electron is involved in the formation of an additional σ-bond. In case of phosphorene, the situation is different. In contrast to atomically flat graphene, phosphorene can be represented as a corrugated monolayer or chains of covalently bonded phosphorus atoms lying in two different planes. Phosphorus atoms in a chain from one plane are covalently bonded with the atoms belonging to the other plane, thus forming the phosphorene layer (Fig. 1a). Phosphorus atoms have five electrons on $3p$ orbitals giving rise to $sp^3$ hybridization. Three electrons participate in the formation of three covalent σ-bonds with neighboring phosphorus atoms, whereas two other electrons occupy a lone pair orbital oriented out-of-plane (Fig. 1a). Phosphorus-phosphorus distance within the planes is 2.23 Å with the P-P-P angle of about 95°, while the values between the planes are 2.29 Å and 102°, respectively.

When monovalent impurities form covalent bond with phosphorus, a redistribution of the electrons between the orbitals takes place (Fig. 1b,c). Now, one σ-orbital form covalent bond with the impurity, two other keep covalent bonds with phosphorus atoms in plane and the last doubly occupied orbital forms a lone pair pointing in the direction of the other phosphorus plane. Let us first consider a situation where phosphorene is fully covered by hydrogen from one side (Fig. 1c). In this case, the covalent bond between the phosphorus planes turns out to be broken, while the distance between the sub-layers increases up to 3.90 Å. Binding energy between the hydrogenated and non-hydrogenated planes of phosphorene amounts to 0.59 eV per P atom that demonstrates relative stability of the interplane bonds. The breaking of covalent bonds induces the $sp^3$ to $sp^2$ transition in the non-hydrogenated plane (Fig. 1c), which transforms itself into phosphorus chains with a slightly decreased P-P distance within the chain (2.20 Å). The appearance of the collectivized π-electrons in single-side hydrogenated phosphorene induces significant changes in the electronic structure (Fig. 2). Particularly, the system becomes metallic with finite density of states at the Fermi level, which is mainly due to the phosphorus atoms in the non-hydrogenated plane.

For double-side hydrogenation, similar rotation of the σ-orbitals occurs as in the case of single-side hydrogenation, but with both layers involved. The planes of hydrogenated phosphorus atoms are now bonded neither by covalent bonds as in pure phosphorene nor by the π-π bonds as in the case of single-side hydrogenation, but by significantly weaker bonds, which is also suggested by the increased interplane P-P distance having a significantly larger value of 3.03 Å and corresponding to a smaller binding energy of 0.39 eV per atom. Obtained values of the interplane binding energies are an order of magnitude higher than in typical dispersive bonds and an order of hydrogen bond strength[43] but few times smaller than the cohesive energy in this material (see Table I). Moreover, as can be seen from Fig.1b, only two σ-orbitals participate in the in-plane bonding and thus each plane is split into weakly bound PH chains aligned along the zigzag direction. The described weakening of the interplane bonds makes the sliding of the planes possible in the horizontal direction, whereas weak interchain bonds could facilitate the movement of chains relative to each other. Finally, it is natural to assume that the weak interaction between the chains makes the entire structure more sensitive to external factors, and could even lead to gradual desorption of individual chains. We believe that flexible thermally fluctuating chains may exist after the phosphorene decomposition due to relatively strong chemical bonding between the atoms in the chain. We note that similar mechanism of phosphorene decomposition in the presence of impurities has been proposed in Ref. 23 designated as "chemical scissors". In order to address this point in more detail, we provide a stability analysis of the fully passivated structures below.

The discussion on hydrogenated phosphorene can be generalized to the case of fluorination. Indeed, we obtain qualitatively the same structures when considering full single-side and double-side fluorination. In this case, however, binding energies are 0.48 and 0.40 eV per atom and interplane distances is 2.51 and 2.64 Å for the cases of single- and double-side fluorination. As one can see, both the interplane and interchain distances are appreciable smaller in the case of fluorine (see Table I), which suggests considerably stronger interactions between both the planes and chains, and thus largely contributes to the overall stability of the structure. This finding is not surprising in view of the additional

2*p* electrons on fluorine, which provide a sizable contribution particularly to the van der Waals interactions.

It is also interesting to note the possibility of existence of different metastable structures with respect to the alignment of hydrogen (fluorine) atoms, which affects the interchain distances. In Table I, we report structural parameters and relative energies of the two structures obtained by imposing different symmetry constraints. If the impurity atoms are restricted to preserve the symmetry of phosphorene, the resulting structure (denoted as "II" in Table I) is somewhat higher in energy than the structure obtained without symmetry constraints ("I" in Table I). This can be understood in terms of an additional electrostatic repulsion between the chains in the former case, which is due to the inability of impurity atoms to fully minimize the overlap between the electron densities. We note that the possibility of other metastable structures, corresponding to different chain stackings is not excluded.

From the point of view of electronic structure, both obtained structures are similar and represent indirect gap semiconductors with the gap of 2.3 – 2.6 (1.8 – 2.3) eV for hydrogenated (fluorinated) phosphorene. In contrast to pristine phosphorene, which is a direct gap semiconductor,[11,12] the band edges are located close to the X and Y points and to X and Γ points, respectively for the valence and conduction bands of the hydrogenated and fluorinated structures (see Fig. 3). The presence of appreciable band gaps, larger than those for pristine mono- and a few layer phosphorene[10-12], suggests relative stability of the obtained structures with respect to further chemical transformations of the hydrogenated (fluorinated) chains. We note that this observation is in contrast with the reported electronic properties of hydrogenated *blue* phosphorus, which is metallic[44] and thus expected to be more reactive. We, therefore, exclude possibility of black phosphorus – blue phosphorus transformation during the hydrogenation (fluorination) of phosphorene.

We now examine dynamical stability of the obtained hydrogenated and fluorinated phosphorene structures. To this end, we calculate phonon spectra, which are shown in Fig. 4. As one can see, while the spectra for both types of structures ("I" and "II") display similar features in the optical part, the spectra corresponding to the metastable structures (PH-II and PF-II) exhibit apparent soft modes along all the directions of the Brillouin zone (marked by dashed lines), pointing to the entire instability of those structures. Similar behavior, though less pronounced, can be seen for the ground state structure of hydrogenated phosphorene (PH-I). In this case, soft modes result predominantly from the out-of-plane vibrations of phosphorus atoms (as can be inferred from the partial density of states in Fig.4a), whereas imaginary frequencies appear along the S-Y direction of the Brillouin zone only. This result allows us to conclude that hydrogenation of phosphorene does not likely lead to dynamically stable 2D structures. Nevertheless, the absence of imaginary frequencies in the Γ-Y direction does not exclude stability of the structure in the direction of the chains. We expect, therefore, that upon full hydrogenation, the whole structure tend to transform into individual loosely bound PH-chains, despite the presence of considerable interchain interactions, as discussed above. Such behavior can be regarded as one of the possible mechanisms leading to phosphorene structural instability in the presence of hydrogen impurities. Qualitatively different picture is observed for the ground state structure of fluorinated phosphorene (PF-I). In this case, soft vibrational modes do not appear in the phonon spectrum, suggesting a dynamically stable configuration. Different stability behavior of the fully hydrogenated and fluorinated phosphorene can be attributed to essentially stronger interchain bonding in the latter case, as discussed above.

**4. Atom-by-atom functionalization of phosphorene by hydrogen and fluorine impurities**

The next step of our study is a step-by-step modeling of single and double-side hydrogenation and fluorination of phosphorene. We first consider the case of atom-by-atom adsorption of hydrogen. In Fig.5a, we show the formation energies as a function of hydrogen coverage. One can see that the formation energy curves are rather oscillating. At low coverages, such behavior can be explained by the unfavorability of the adsorption of an odd number of hydrogen atoms. Indeed, when single hydrogen is chemisorbed, the covalent bond between the planes breaks, being a consequence of the $sp^3$ to $sp^2$ transition on the adjacent phosphorus atom belonging to the other plane (that is on the second nearest neighbor) as shown in Fig. 1c. At the same time, an electron on the $p_z$-orbital of this atom becomes unpaired without forming π-like bonds. The most energetically favorable second adsorption site corresponds to the third nearest-neighbor phosphorus atom that is the nearest atom along the zigzag direction. This configuration is about 0.3 eV lower in energy than those with the second hydrogen adsorbed on the first or second nearest-neighbor phosphorus atom. The adsorption of second hydrogen also results in a $sp^3$ to $sp^2$ transition on the second nearest-neighbor with the appearance of a second $p_z$-orbital, giving rise to an in-plane π-bond, lowering the total energy of the system. The subsequent steps of hydrogenation follow similar scenarios until the entire row of phosphorus atoms is filled by hydrogens. At the following steps of hydrogenation, other rows begin fill in the similar manner, depending on the possibility of adsorption from one or both sides. Therefore, both single- and double-side hydrogenation can be considered as a row-by-row functionalization process. As can be seen from Fig.5, single-side adsorption is less favorable, which can be attributed to the existence of weakly bound π-electrons in the non-passivated plane of phosphorene, whereas upon double-side hydrogenation, π-electrons are available to form strong σ-bonds with hydrogen atoms. It is important to note that double-side hydrogenation results in a number of metastable $P_xH_y$ structures, in which hydrogen adsorbed along the rows as shown in the insets of Fig. 5.

In contrast to hydrogenation, fluorination of phosphorene is always an exothermic process, which can be understood in terms of significantly stronger fluorine-phosphorus bonds. In this case, the formation energy curves shown in Fig. 5b follow the same steps of adsorption as for the hydrogenation. This observation is not surprising in view of equal valence of both atoms.

**5. Molecular dissociation and migration of hydrogen and fluorine on phosphorene**

We now discuss the possibility of adsorption of hydrogen and fluorine from molecular gas phase. To get insight into the energetic favorability of such a process, it is important to take energy barriers resulting from the molecular dissociation into account. To this end, we consider three possible final states of the dissociation of a diatomic molecule on phosphorene as shown in Fig.6: (a) *along* the chains, (b) *across* the chains, and (c) as a *bridge* between the different chains. For these states, we perform the calculation of energy barriers by scanning the whole potential energy landscape and identifying the minimum energy path between the initial and final states of hydrogen and fluorine. As the initial state, we consider free molecules placed at a considerable vertical distance from the surface above the desired adsorption sites. The corresponding minimum energy paths and energy barriers for dissociation are shown in Fig.5d,e. One can see that the energy barriers depend strongly on the particular molecule and the final adsorption site. In case of $H_2$, the lowest energy barrier corresponds to the dissociation across the phosphorene chain (Fig.6b) with a rather high value of 2.54 eV, similar to graphene.[45] The most energetically favorable scenario for the functionalization by atomic hydrogen (along the chains, Fig.6a) corresponds to the highest value of the energy barrier (3.68 eV). In the case of $F_2$, the dissociation barriers are significantly lower for all three scenarios and do not exceed 1 eV, which is due to high chemical activity of fluorine. The lowest energy barrier (0.19 eV) also corresponds to the scenario different from the case of atom-by-atom fluorination that is to the bridge configuration (Fig.6c). In the case of functionalization by monovalent species, the most favorable adsorption sites are primarily determined by the electronic structure of phosphorene and the adsorbates, while the difference between the adsorbates is only significant for the determination of the formation energies (Fig. 5a,b). In case of molecular dissociation, another factors come into play. In particular, a crucial role plays the mismatch between the interatomic distances in the molecule and P-P distance in phosphorene. In the case of $H_2$, with an interatomic distance of 0.74 Å, it is natural to expect the lowest energy barrier for the dissociation on the nearest atoms in phosphorene. On the other hand, the F-F distance in $F_2$ is significantly larger (1.42 Å), which favors the dissociation on more distant phosphorus atoms.

In order to assess the possibility of diffusion of the adsorbed atoms on the surface of phosphorene toward the most energetically favorable configurations (i.e., along the chain), we calculate the migration barriers by considering the minimum energy path of a single atom. The results presented in Fig.5f indicate that for hydrogen, the migration barrier is significantly smaller than the barrier of dissociation (1.09 vs 2.54 eV), while for fluorine the opposite situation takes place (2.81 vs 0.19 eV). The difference between hydrogen and fluorine can be described not only by the stronger P-F bonds but also by different the migration pathways. Hydrogen migrates over P-P bonds in the vicinity of electrons occupying the rather diffusive lone pairs of phosphorene, which provide additional repulsion along the pathway, making the transition state of hydrogen particularly unfavorable. In contrast, migration of fluorine is taking place between the prosphorene chains, which is more favorable due to the reduced repulsion between the valence electrons of fluorine and neighboring occupied phosphorus orbitals along the bridge pathway.

One can see that the total energy barrier between the molecular state and the most favorable atomic configurations, determined as the highest value between the dissociation and migration barriers, is rather high for either hydrogen or fluorine (2.54 and 2.81 eV, respectively). We conclude, therefore, that the formation of regular row-like hydrogen or fluorine patterns, resulting from the atom-by-atom adsorption, is unlikely to achieve from molecular gas phase. On the other hand, the dissociation of fluorine is significantly less energetically demanding process compared to hydrogen and, therefore, passivation of phosphorene by fluorine molecules expected to be more efficient in practice.

**6. Oxidation of phosphorene**

In contrast to the case of hydrogenation and fluorination, chemisorption of oxygen atoms does not necessarily require breaking of covalent bonds. Both electrons of the lone pair belonging to a phosphorus atom (Fig. 1a) participate in the formation of a double bond with oxygen. This observation is in line with the result presented in Fig.5c, displaying that the oxidation of phosphorene is an exothermic process, whose efficiency does not depend significantly on the number of oxidized sides. The fluctuating behavior of the formation energy curves, similar to the cases of hydrogen and fluorine, is noticeable at initial steps when the perturbation of phosphorene caused by oxidation is rather large. However, when symmetry of phosphorene is broken by the chemisorption of a few first oxygen atoms, further oxidation does not provide such significant distortions. Recent work reported rather low (0.54 eV) value of the energy barrier for the dissociation of molecular oxygen on phosphorene.[20] This value is about two times smaller than reported for graphene,[46] which suggests that spontaneous dissociation of oxygen molecule on phosphorene is not excluded. As has been recently demonstrated, dynamically stable regular structures might appear at low concentrations of oxygen impurities.[25] At high concentrations, however, our results suggest that the formation of regularly ordered phosphorene-oxygen structures is not energetically favorable, which is different from the case of monovalent impurities. To interpret this, we note that in addition to the double bonds with a single phosphorus atom, oxygen can also form two single bonds with two phosphorus atoms. Indeed, as can be seen from the insets in Fig.5c, at a high level of single- or double-side oxidation, the formation of the described type of bonds takes place and the process of uniform oxidation turn into the formation of disordered $P_xO_y$ systems. Significant energy gain at the first steps of the disordered phosphorene oxide formation makes this amorphous-like structure more energetically favorable with respect to the ordered $P_xO_y$ layer. Thus we can conclude that the oxidation of phosphorene likely leads to the formation of a disordered P-O phase instead of an ordered phosphorene oxide discussed in previous theoretical works.[25,33]

## 7. Conclusions

In summary, based on first-principles calculations, we demonstrate that single-side atom-by-atom coverage of phosphorene with monovalent impurities such as hydrogen or fluorine results in a transition from $sp^3$ to $sp^2$ hybridization of phosphorus atoms and make the resulting structures metallic. Double-side coverage of phosphorene with hydrogen or fluorine monoatomic impurities can be a source of various $P_xH_y$ ($P_xF_y$) metastable structures where impurities form regular 1D rows on phosphorene. Although such regular patterns are the most favorable energetically, it appears unlikely to achieve them by dissociating the $H_2$ or $F_2$ molecules from gas phase, which is due to high energy barriers (>2.5 eV), determined by the dissociation itself and by migration of atoms. Full hydrogenation or fluorination of phosphorene leads to a decomposition of its structure into narrow PH (PF) chains bonded by hydrogen-like bonds. From the electronic structure of view, both systems correspond to indirect gap semiconductors with an energy gap of ~2.3 eV. In contrast to the fluorinated structure, which proven to be dynamically stable, fully hydrogenated phosphorene is unstable due to significantly weaker interchain coupling, resulting into the appearance of soft vibrational modes. Given that the adsorption of monoatomic fluorine is always an exothermic process and the dissociation from molecular gas phase is relatively efficient due to low energy barriers (0.19 eV), fluorinated phosphorene structures are expected to be easily achievable in practice and, therefore, can be considered as a testbed for experiments. Finally, both single- and double-side atom-by-atom oxidation of phosphorene also proven to be exothermic processes, whereas moderate energy barriers for dissociation (~0.54 eV[20]) make spontaneous oxidation possible. In contrast to monovalent impurities, in case of oxygen we found a transition to a disordered amorphous-like 2D system after oxidation of more than half of the surface.

**Acknowledgements**

The authors acknowledge support from the European Union Seventh Framework Programme under Grant Agreement No. 604391 Graphene Flagship and from the Ministry of Education and Science of the Russian Federation, Project N 16.1751.2014/K.

**Tables and figures**

**Table I.** Total energies relative to the ground state, structural parameters and *GW* energy gaps for two structural configurations (denoted I and II) of fully hydrogenated phosphorene (PH) presented in Fig. 1b. In parenthesis, the same values for fluorinated phosphorene (PF) are given.

| Structure | PH-I (PF-I) | PH-II (PF-II) |
|---|---|---|
| Total Energy, meV/P | 0 (0) | 34.9 (13.8) |
| *a*/*b* | 0.76 (0.83) | 0.76 (0.83) |
| Distance between planes, Å | 2.87 (2.19) | 3.41 (2.40) |
| Distance between chains, Å | 2.88 (2.64) | 2.83 (2.60) |
| Point symmetry group | $C_{1h}$ | $D_{2h}$ |
| Energy gap, eV | 2.29 (2.27) | 2.64 (1.82) |

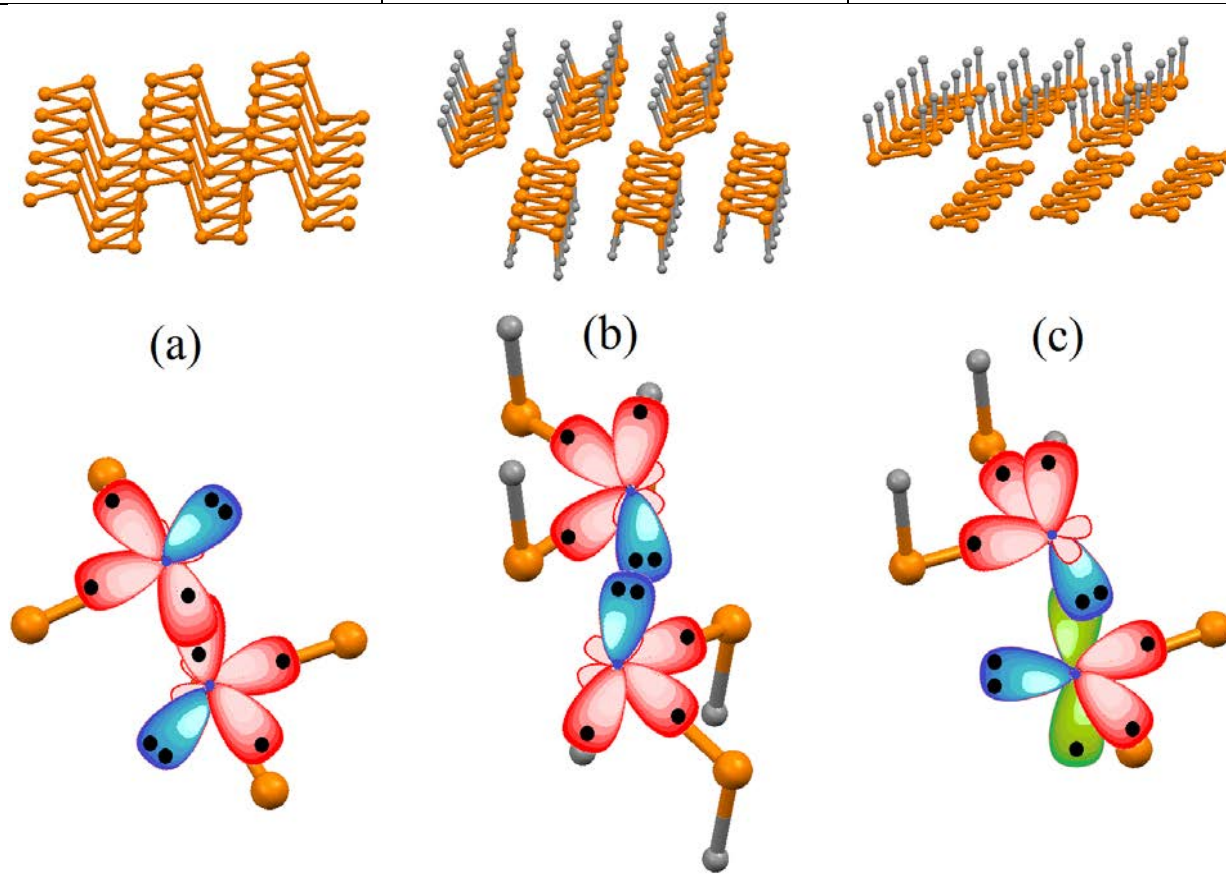

**Figure 1.** Optimized atomic structures and sketches of *sp*-hybridization of phosphorus orbitals for (a) pure, and hydrogenated from (b) both and (c) single sides phosphorene. σ-orbitals, lone pairs and $p_z$-orbitals are shown by red, blue, and green, respectively.

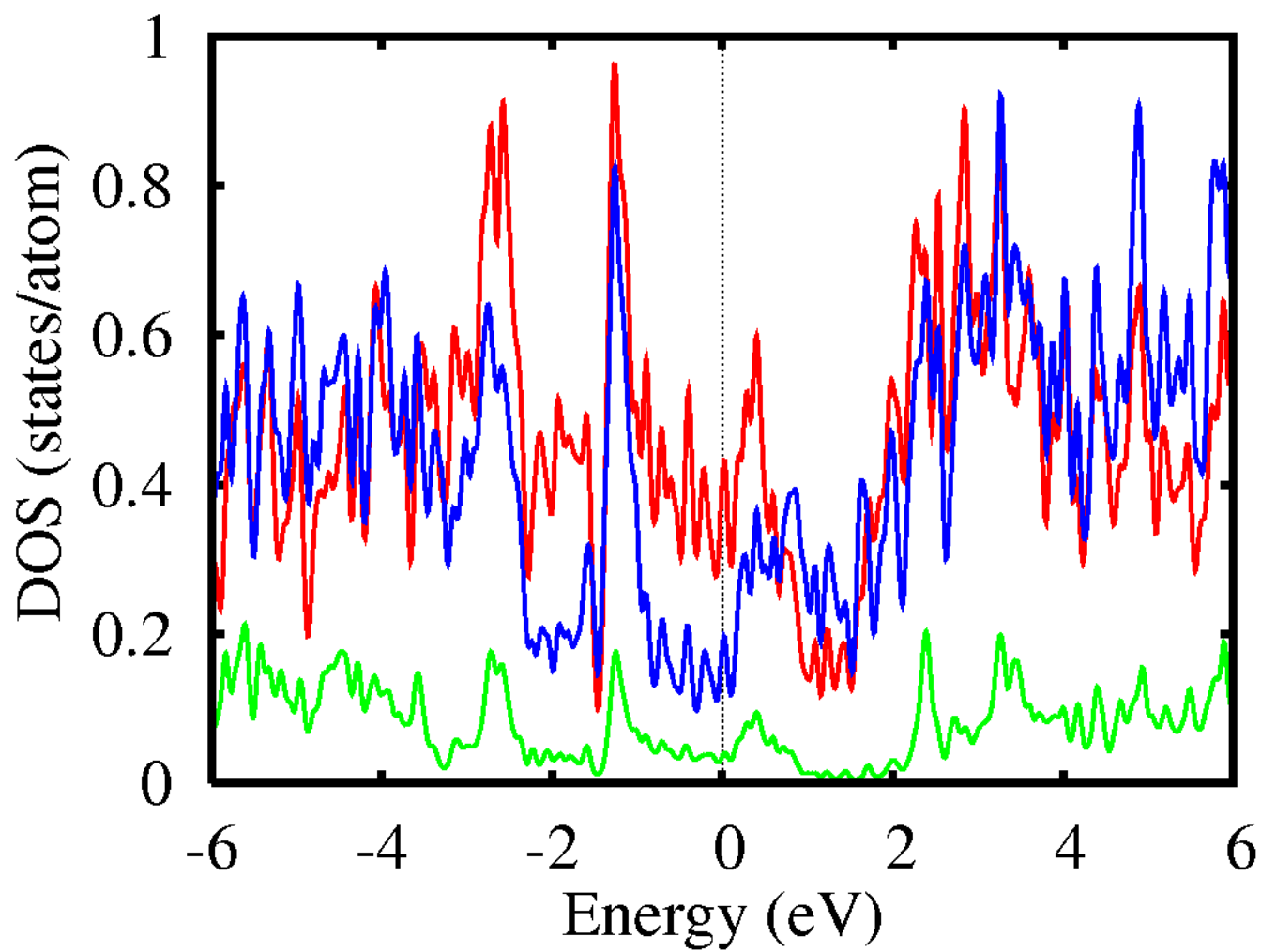

**Figure 2.** Densities of states projected on non-hydrogenated (red) and hydrogenared (blue) phosphorus atoms as well as on hydrogen atoms (green) in single-side fully hydrogenated phosphorene (Fig. 1c).

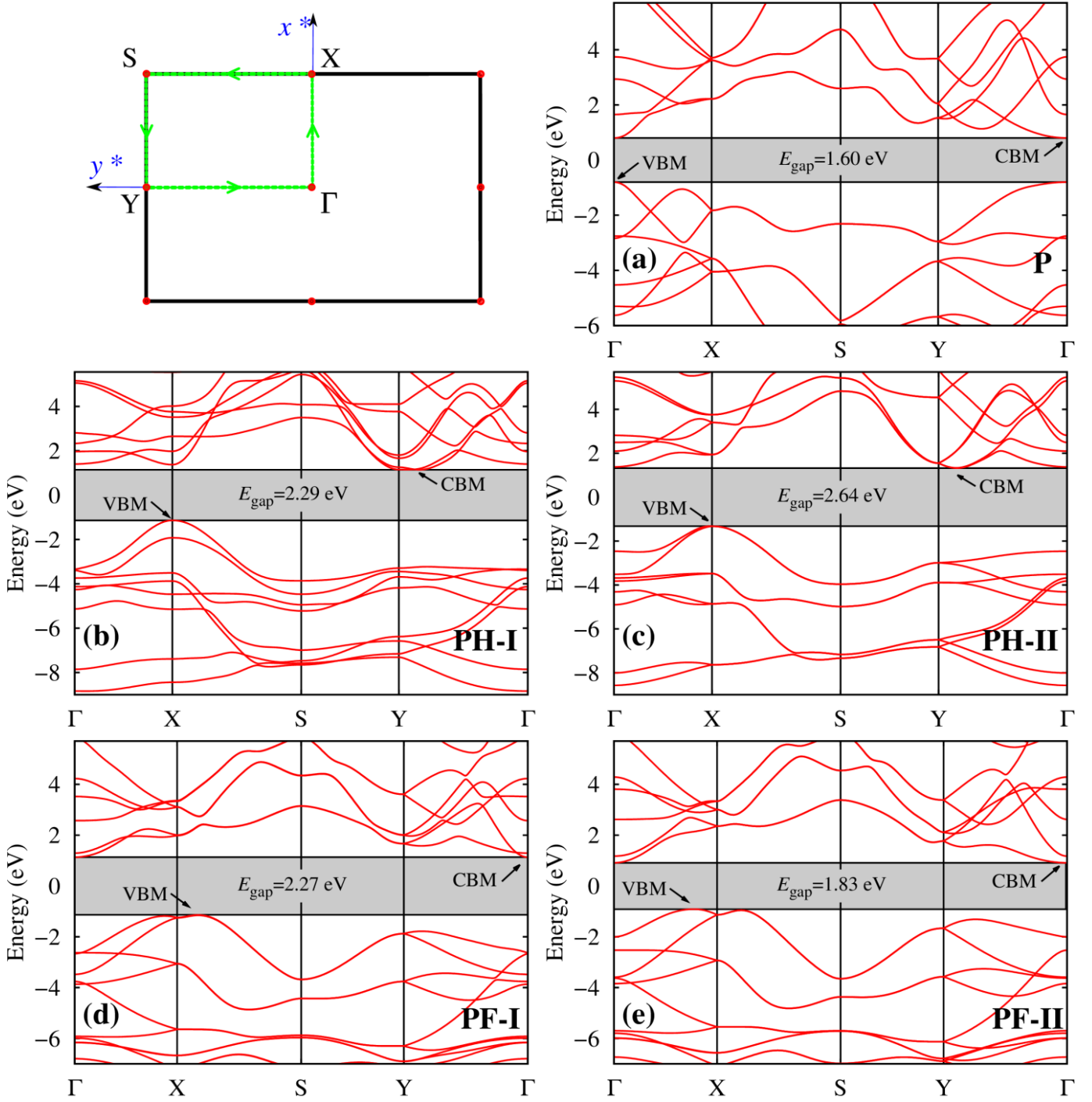

**Figure 3.** Band structures calculated by the *GW* method for (a) pristine phosphorene and two structural configurations of fully hydrogenated (b,c) and fluorinated (d,e) phosphorene obtained in this work (see text for details). In the upper left corner, the Brillouin zone and the corresponding path along the high-symmetry points are shown. VBM and CBM stand for the valence band maximum and conduction band minimum, respectively.

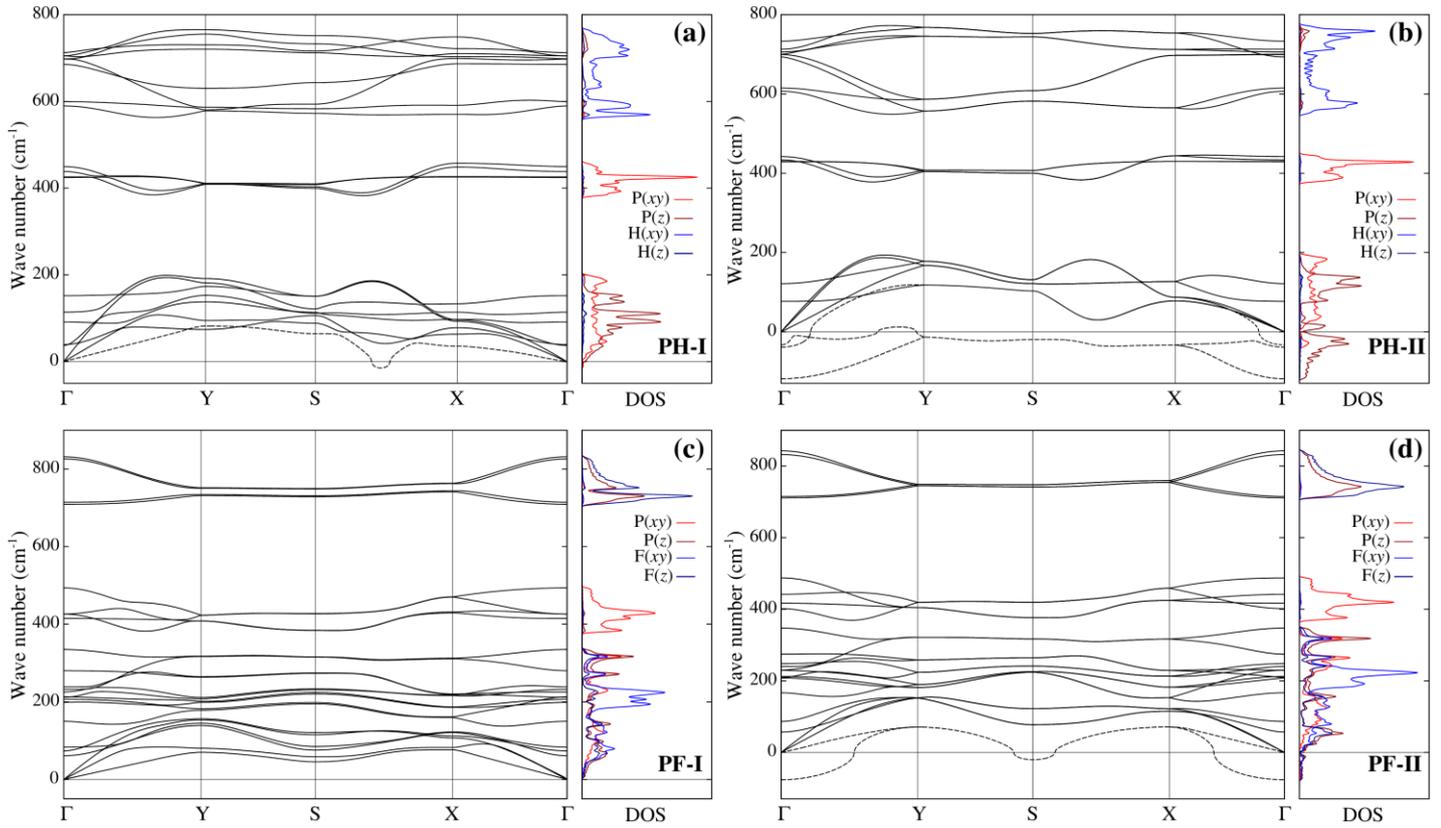

**Figure 4.** Phonon dispersion relation and partial density of states (DOS) of the two structural configurations of fully hydrogenated (a,b) and fluorinated (c,d) phosphorene. The partial DOS are projected on in-plane (*xy*) and out-of-plane (*z*) modes associated with the vibrations of P and H (F) atoms. Dashes lines correspond to soft modes with imaginary frequencies (represented as negative values of the wave number).

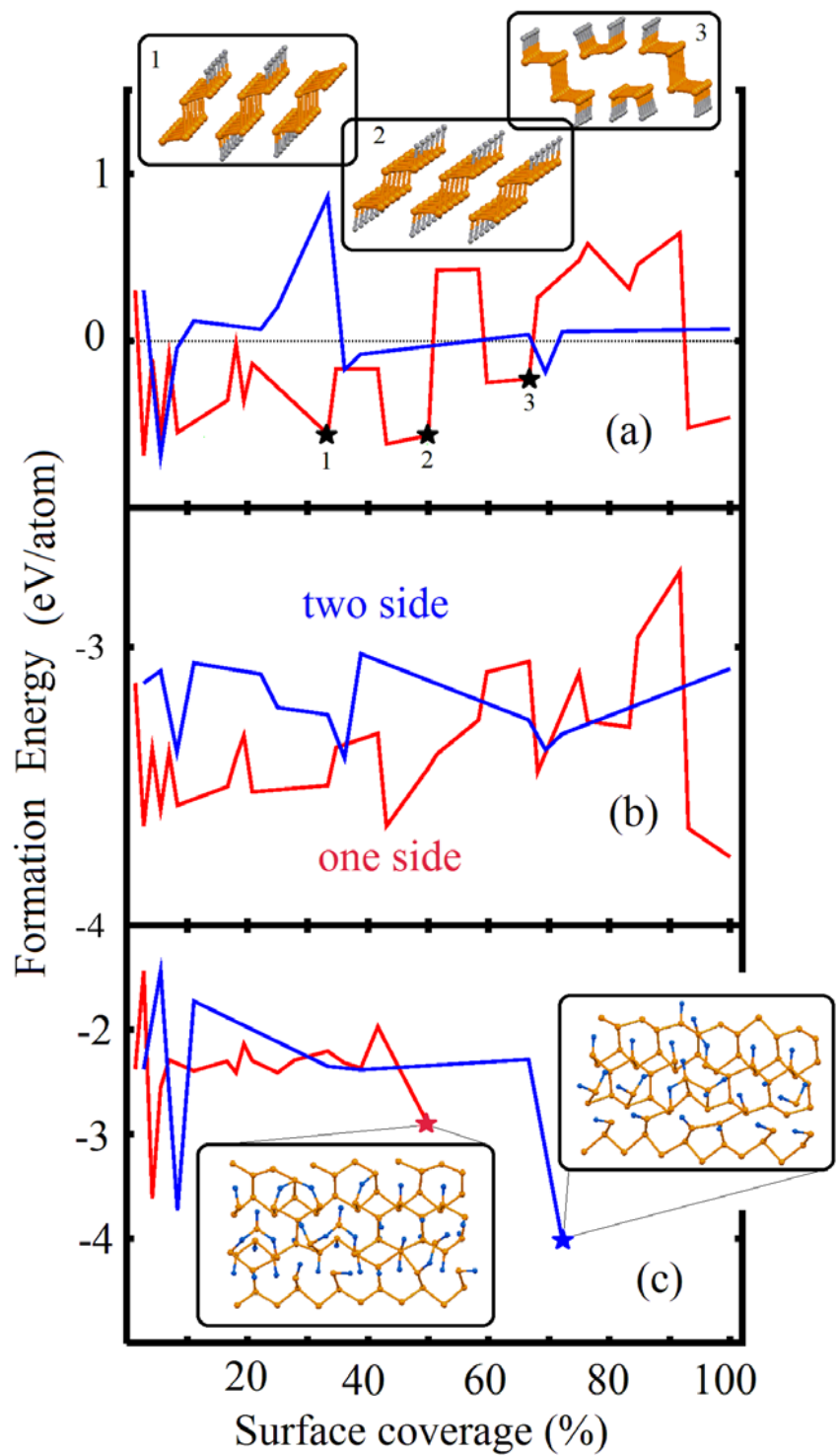

**Figure 5.** Formation energies for single- and double-side hydrogenation (a) fluorination (b) and oxidation (c) of phosphorene. In the insets of panel (a), optimized atomic structures of metastable intermediate configurations corresponding to the local formation energy minima (indicated by stars) are present. In the insets of panel (c), optimized atomic structures for the first steps of the disordered phosphorene oxide formation are shown. Surface coverage defined as percentage of functionalized atoms available for this process in each case.

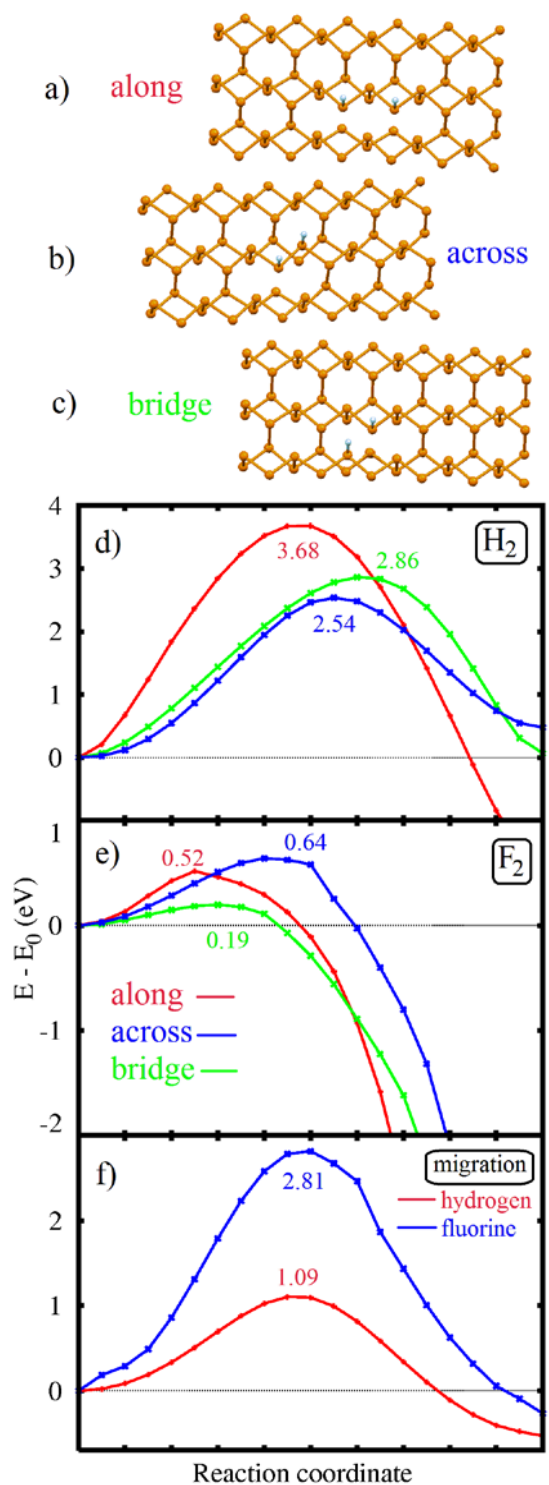

**Figure 6.** (a-c) Optimized atomic structure of the three possible final states of dissociation of $H_2$ and $F_2$ molecules on phosphorene. (d,e) Minimum energy paths calculated for the dissociation of $H_2$ and $F_2$ molecules, respectively, for different final adsorption sites. (f) Minimum energy paths for migrations of hydrogen and fluorine atoms from the configuration with the lowest barrier of dissociation (according to (d) and (e)) to the most energetically favorable configurations (i.e., *along* the chain, as shown in (a)). Values indicate the corresponding energy barriers.